\documentclass[reprint, aps,pra,twocolumn,showpacs,amsmath,amssymb, floatfix]{revtex4-2}
\usepackage{graphicx}
\usepackage{dcolumn}
\usepackage{bm}
\usepackage{bbm}
\usepackage{amssymb}
\usepackage[table]{xcolor}
\usepackage{colortbl}
\usepackage{lipsum}
\usepackage{braket}
\usepackage{multirow}
\usepackage[colorlinks=true,linkcolor=blue, citecolor=red, urlcolor=magenta]{hyperref}
\usepackage{url}
\begin{document}

\preprint{APS/123-QED}

\title{
SU(2) gadget for higher-order Poincar\'{e} sphere
}
\author{Mohammad Umar}
 \email{aliphysics110@gmail.com}{}
\author{Paramasivam Senthilkumaran}
 \email{psenthilk@yahoo.com}
\affiliation{
Optics and Photonics Centre\\
Indian Institute of Technology Delhi\\New Delhi 110016, INDIA
}

\begin{abstract}
The combination of two quarter-wave plates and one half-wave plate, regardless of their sequential arrangement, constitutes a well-established universal SU(2) gadget capable of implementing all polarization transformations on the standard Poincar\'{e} sphere. However, there is no analogous system for realizing all polarization transformations on the higher-order Poincar\'{e} sphere, a member of a higher topological index space. This work demonstrates that an optical gadget, comprising two quarter-wave $q$-plates and one half-wave $q$-plate, arranged in any order, is an SU(2) gadget to realize arbitrary polarization evolution on the higher-order Poincar\'{e} sphere. 

\end{abstract}

\maketitle

\begin{figure*}[t]
\includegraphics[scale=0.42]{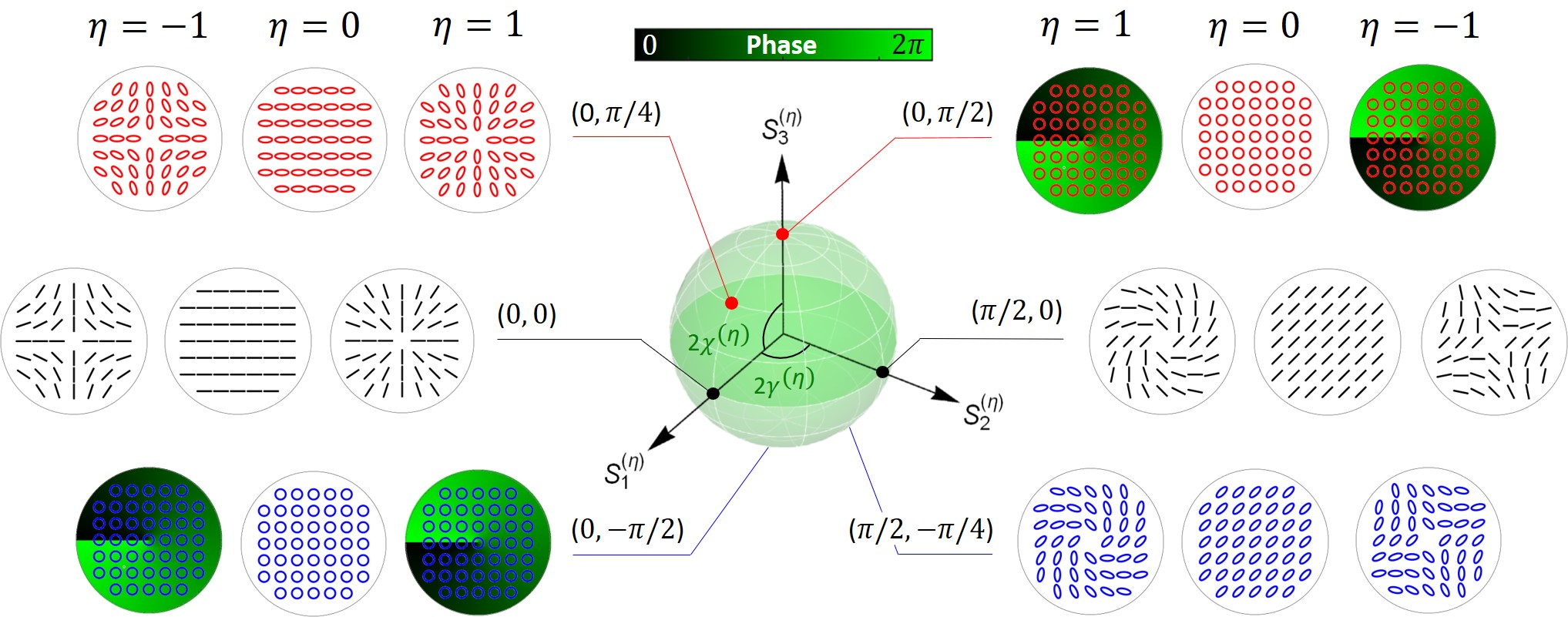}
\caption{\label{ps_01}(Color online). Illustration of the polarization distributions corresponding to the topological spheres of orders $\eta = -1$, $0$ and $+1$. The polarization ellipses are color-coded: red denotes right-handed polarization and blue denotes left-handed polarization. The sphere coordinates are defined as ($2\gamma^{(\eta)}$, $2\chi^{(\eta)}$). The phase distribution of the eigenstates is shown in the background for the two orthogonal polarization basis states.}
\end{figure*}

The Poincar\'{e} sphere (PS), an $\mathbf{S}^{2}$ sphere \cite{poincare1954theorie},
provides a geometric framework to represent the polarization of light, where all the polarization states are mapped to its surface through the Stokes parameters as Cartesian
coordinates (Fig. \ref{ps_01}) \cite{stokes1851composition}. Each point on the PS corresponds to a homogeneous state of polarization (SOP) distribution,
with transformations between them described by elements of the three-parameter group
called the SU(2) group. The SU(2) transformations between the SOPs are equivalent to rotations on the PS governed by the SO(3) group, reflecting the two-to-one homomorphism between SU(2) and SO(3) groups. Birefringent media, such as waveplates, serve as a practical example of SU(2) elements in this context. Before proceeding to the main theme of the paper, we provide a brief introduction to the topological sphere and structured optical elements.\\
\indent
Waveplates, such as quarter-wave plates (QWPs) and half-wave plates (HWPs), exemplify SU(2) elements that perform rotations of $\pi/2$ and $\pi$ on the PS, respectively, about specific axes decided by the fast axis orientation of the waveplate. However, to perform arbitrary rotations around arbitrary axes on the PS, a more general SU(2) element is
 required. Motivated by this, R. Simon et al. constructed SU(2) gadgets, comprising of finite number of standard waveplates, capable of realizing polarization transformations on the PS \cite{simon1989hamilton, simon1989universal, simon1990minimal}. Finally, they introduced a minimal SU(2) gadget, consisting of two QWPs and one HWP, arranged in any order (HQQ, QHQ, QQH), which can realize all possible SU(2) transformations on the PS \cite{simon1990minimal}. Here Q denotes QWP and H denotes HWP. This configuration is now widely recognized as a well-established \textit{universal} SU(2) gadget that enables transportation of a state $A$ to any arbitrary state $B$ on the PS.\\
 \indent
Despite the well-celebrated geometry of the PS, it is inadequate for representing higher-order solutions of Maxwell’s equations that admit spatially inhomogeneous beams. These beams form connected regions on the PS rather than single points. To circumvent this, G. Milione et al. introduced the higher-order Poincaré sphere (HOPS), where inhomogeneous beams are represented as points on its surface \cite{milione2011higher}. Like the PS, HOPS is also an $\mathbf{S}^{2}$ sphere and homomorphic to it. The PS represents spin polarization states, serving as its foundational basis. In contrast, HOPS advance this concept by incorporating both spin angular momentum (SAM) and orbital angular momentum (OAM), thereby representing polarized singular beams with a uniform ellipticity throughout their cross-sectional profile.

The HOPS beams are the superposition of right- and left-circularly polarized vortex beams with opposite topological charges \cite{ruchi2020phase}. The inhomogenous polarization distribution of HOPS beam has constant ellipticity $\chi$, but spatially varying polrization azimuth $\gamma$, and is characterized by the Poincar\'{e}-Hopf (PH) index \cite{freund2002polarization} expressed as $\eta =(1/2\pi)\oint \nabla \gamma\cdot dl$. The line integral is computed over a closed contour surrounding the singularity, where the azimuth becomes singular or undefined. The sign of the PH index characterizes the rotational sense of the azimuthal variation of the SOPs in the vicinity of the singularity. In this view, the PH index not only defines the HOPS beam but also quantifies the order of the HOPS and accordingly, the beam is structured. Mathematically, these beams are expressed as
 \begin{equation}
     \ket{\psi_{\ell}}= \psi_{R}\ket{R_{\ell}}+\psi_{L}\ket{L_{\ell}}.
     \label{psi_01}
 \end{equation}
Here, the basis states, defined as $\ket{R_{\ell}}=e^{-i\ell\phi}(\hat{\textbf{x}}-i\hat{\textbf{y}})/\sqrt{2}$ and $\ket{L_{\ell}}=e^{i\ell\phi}(\hat{\textbf{x}}+i\hat{\textbf{y}})/\sqrt{2}$, representing the right circular polarized (RCP) and left circular polarized (LCP) optical vortex (OV) of topological charge $-\ell$ and $\ell$ respectively and $\psi_{R}$ and $\psi_{L}$ represent the corresponding complex amplitudes respectively. For $\ell=0$, the eigenstates represent plane waves and the corresponding beam is homogeneously polarized. The PH index for the HOPS beam is $\eta=\ell$. For $\ell\ge 1$ and $\ell\le -1$, the HOPS beam contains the polarization singularity known as the V-point singularity \cite{ruchi2020phase}. The polarization distributions corresponding to the HOPS of orders $\eta=-1$, $0$ and $+1$ are illustrated in Fig. \ref{ps_01}. The Stokes parameters (SPs) associated with the field given in Eq. (\ref{psi_01}) is expressed as $S_{0}^{(\eta)} = |\psi_R|^{2} + |\psi_L|^{2}$, $S_{1}^{(\eta)} = 2\texttt{Re}[\psi_{R}\psi_{L}^{*}]$, $S_{2}^{(\eta)} = 2\texttt{Im}[\psi_{R}\psi_{L}^{*}]$ and $S_{3}^{(\eta)} = |\psi_R|^{2} - |\psi_L|^{2}$. These SPs are used to craft HOPS as shown in Fig. \ref{ps_01}. The higher-order SPs satisfying the relation $(S_{0}^{(\eta)})^2 = (S_{1}^{(\eta)})^2 + (S_{2}^{(\eta)})^2 + (S_{3}^{(\eta)})^2$. The coordinates, longitude $2\gamma^{(\eta)}$ and latitude $2\chi^{(\eta)}$, of the HOPS are given by $2\gamma^{(\eta)}=\tan^{-1}(S_{2}^{(\eta)}/S_{1}^{(\eta)})$ and $2\chi^{(\eta)}=\sin^{-1}(S_{3}^{(\eta)}/S_{0}^{(\eta)})$, respectively. For $\eta=0$, the new SPs reduce to standard plane wave SPs.\\
\indent
The configurations HQQ, QHQ and QQH are the universal SU(2) gadgets to realize complete coverage on PS. However, no such type of SU(2) gadget is available to realize complete coverage of the HOPS, and this paper addresses this gap. Recently, an optical gadget has been proposed in which two HWPs are placed between $q^{Q}$-plates to achieve arbitrary polarization transformations on the HOPS, where successive non-holonomic polarization transformations govern the polarization evolution, making it not an SU(2) gadget \cite{bansal2025gadget}. Our proposed SU(2) gadget consists of two $q^{Q}$-plates and one $q^{H}$-plate, with the condition that all the $q$-plates have the same topological charge, arranged in any order. This configuration is capable of providing continuous holonomic polarization transformations on the HOPS.\\
\begin{figure}[b]
\centering
\includegraphics[scale=0.21]{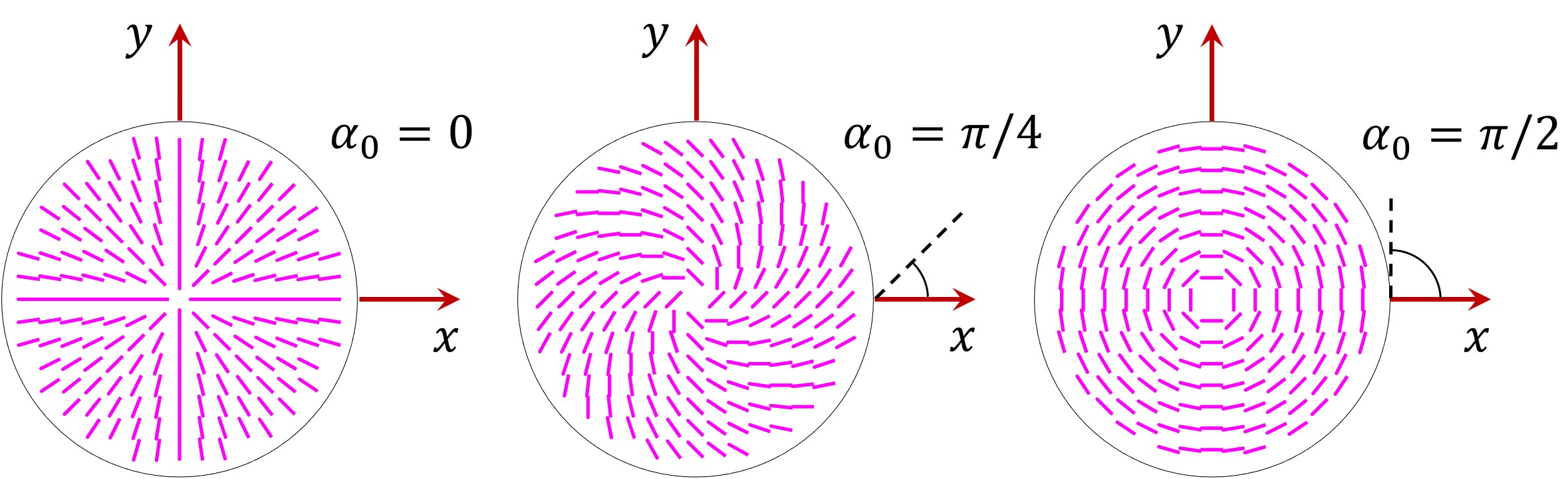}
\caption{\label{fig:ps_006} (Color online). The geometry of the $q$-plate with topological charge $q = 1$ is shown for different offset angles: $\alpha_{0} = 0$ (first structure), $\alpha_{0} = \pi/4$ (second structure) and $\alpha_{0} = \pi/2$ (third structure).}
\end{figure}
\indent
A $q$-plate is a birefringent element \cite{marrucci2006optical, marrucci2013q, rubano2019q, delaney2017arithmetic, machavariani2008spatially, kadiri2019wavelength, bansal2020use, cardano2012polarization, slussarenko2011tunable}, a member of the SU(2) family, where the fast axis orientation is spatially varying. The fast axis orientation is given as $\alpha(\phi)=q\phi+\alpha_{0}$, where $q$ is the topological charge and is given by $q=(1/2\pi)\oint\nabla \alpha(\phi)\cdot dl$ and $\alpha_{0}$ is the offset angle which represents the fast axis orientation with respect to a given reference axis. According to this definition, $q$-plates with a given topological charge $q$ represent the entire class of retarders characterized by a retardance $\delta$. A $q$-plate with a retardance of $\pi$ is referred to as a half-wave $q$-plate ($q^H$-plate), while one with a retardance of $\pi/2$ is known as a quarter-wave $q$-plate ($q^Q$-plate). Fig. \ref{fig:ps_006} presents the geometry of the $q$-plate with $q = 1$ for offset angles $\alpha_{0} = 0$, $\pi/4$ and $\pi/2$. The SU(2) Jones matrix of a waveplate having retardance $\delta$ and fast axis orientation $\alpha$ is given by \cite{collett2005field}
\begin{equation}
J_{\alpha}= 
\begin{bmatrix}
        \cos \frac{\delta}{2} + i\sin\frac{\delta}{2} \cos 2 \alpha & i\sin\frac{\delta}{2} \sin 2 \alpha \\[10pt]
        i\sin\frac{\delta}{2} \sin 2 \alpha & \cos \frac{\delta}{2} - i\sin\frac{\delta}{2} \cos 2 \alpha
\end{bmatrix}.
\label{matrix01}
\end{equation}
\noindent
By substituting the fast axis orientation $\alpha \rightarrow \alpha(\phi)$ in the given matrix, the resulting expression corresponds to the Jones matrix of a $q$-plate. The above matrix is a symmetric matrix.\\
\indent
The HOPS beam and the $q$-plate share the same topological feature, as one of their defining parameter is azimuthally varying. Levergaing this, we have shown that a \textit{general} $q$-plate, under the holonomy condition $q=\eta$, is a suitable anisotropic structured element to travel on the surface of HOPS \cite{umar20252, umar2025holonomically}. An SU(2) transformation on the HOPS, mediated by the $q$-plate, results in a one-to-one mapping between each SOP of the input beam and the corresponding SOP of the output beam. In that case, a single \textit{global} SO(3) rotation on the HOPS is the collection of many \textit{local} SO(3) rotations on the standard PS. \\
\indent
Again, leveraging the fact that HOPS beam and the $q$-plate share the same topological textures, the concept of topological index spaces has been introduced recently \cite{umar2025holonomically} to address polarization-structured electromagnetic fields and birefringent elements with engineered anisotropies in specific topologies. Each family of the index space comprises two types of members: the topological sphere and the corresponding structured elements that can perform holonomic polarization transformations on that sphere. Therefore, the optical gadget presented in \cite{bansal2025gadget} could serve as an example of a mixed index space.\\
\indent
Recently, we have shown that combinations such as $q^{Q}q^{Q}q^{H}$, $q^{Q}q^{H}q^{Q}$ and $q^{H}q^{Q}q^{Q}$, under specific constraints, act effectively as a single $q$-plate and exhibit the feature of tunable retardance ranging from $0$ to $2\pi$, where this feature is controlled by the relative offset angle of the involved plates \cite{umar2025mathematics}. It is shown that the retardance of the $q$-plate represents the rotation angle on the HOPS of order $\eta$, provided that the holonomy condition is met \cite{umar20252}. Therefore, these configurations allow a complete $2\pi$ rotation on the HOPS, with the rotation axis governed by the offset angle of the resultant $q$-plate. This paper is motivated by these findings. \\
\indent
In the canonical parameterization, the SU(2) element is expressed as \cite{simon1989universal}
\begin{equation}
    U(\psi, \theta, \phi)=e^{\pm i\frac{\Omega}{2} \mathbf{n}(\theta, \phi).\sigma},
\end{equation}
where \textbf{$\sigma$} is the Pauli spin matrices and is given by 
\begin{equation}
\sigma_1 = \begin{bmatrix} 0 & 1 \\ 1 & 0 \end{bmatrix}, \quad
\sigma_2 = \begin{bmatrix} 0 & -i \\ i & 0 \end{bmatrix}, \quad
\sigma_3 = \begin{bmatrix} 1 & 0 \\ 0 & -1 \end{bmatrix}.
\end{equation}
The term, $\mathbf{n}(\theta, \phi)=(\sin \theta \cos \phi, \sin \theta \sin \phi, \cos \theta)$ specifies the axis of rotation in $\textbf{R}^{3}$ space and $\Omega$ is the rotation angle which defines the magnitude of the transformation.\\
\indent 
The derivation of the Simon-Mukunda gadget starts with an Euler angle parameterization with the Euler angles $\xi$, $\rho$ and $\zeta$, as it offers a convenient framework for decomposing SU(2) transformations into sequential rotations. Consider an SU(2) element expressed as
\begin{equation}
U(\xi, \rho, \zeta)=e^{-i\frac{\xi}{2} \sigma_{2}}e^{i\frac{\rho}{2} \sigma_{3}}e^{-i\frac{\zeta}{2} \sigma_{2}}.
\label{euler01}
\end{equation}
Further, it is straightforward to show that \cite{simon1990minimal}
\begin{align}
     U(\xi, \rho, \zeta)& = Q_{\frac{\xi}{2}+\frac{\pi}{4}}H_{-\frac{\pi}{4}+\frac{\xi+\rho-\zeta}{4}}Q_{\frac{\pi-2\xi}{4}}\\
     & = H_{-\frac{\pi}{4}+\frac{\xi+\rho-\zeta}{4}}Q_{\frac{\pi}{4}+\frac{\rho-\zeta}{2}}Q_{\frac{\pi}{4}-\frac{\zeta}{2}} \\
     & = Q_{\frac{\xi}{2}+\frac{\pi}{4}}Q_{\frac{\pi}{4}+\frac{\rho+\zeta}{2}}H_{-\frac{\pi}{4}+\frac{\xi+\rho-\zeta}{4}},
\end{align}
where $Q_{\beta}$ and $H_{\beta}$ represent the Jones matrices of the QWP and HWP, respectively, with their fast axis orientation at $\beta$. This demonstrates that every element of SU(2) (Eq. (\ref{euler01})) can be realized using two QWPs and one HWP, regardless of their sequential arrangement. This is the way that the universal SU(2) gadget for PS is established.\\
\indent
To design the SU(2) gadget for HOPS, again, expressing the parameter in terms of Euler angles proves to be convenient. Here, we \textit{modified} the Euler angle parameterization such that
\begin{equation}
     \mathcal{U}(\xi, \rho, \zeta) =e^{-i\frac{\xi}{2} \sigma_{2}}e^{i\frac{\rho}{2}[(\sin{2q\phi})\sigma_{1}+ (\cos{2q\phi}) \sigma_{3}]}e^{-i\frac{\zeta}{2} \sigma_{2}},
     \label{euler02}
\end{equation}
\\
where $q \in \mathbb{R}$ and $\phi$ run from $0$ to $2\pi$. The given expression reduces to Eq. (\ref{euler01}) when $q=0$. Here, the middle exponential term represents an SO(3) rotation about axes lying in the $xz$-plane, with their orientation controlled by the parameters $q$ and $\phi$. This \textit{modified} parameterization also fully spans the SU(2) group, despite the rotation axis in the second exponential being space-variant. By tuning the Euler angles $\xi$, $\rho$ and $\zeta$, any SU(2) operation can be realized, ensuring that there are no constraints on the realizability of SU(2) element. Further, Eq. (\ref{euler02}) can also be expressed as
\begin{widetext}
\begin{equation}
     \mathcal{U}(\xi, \rho, \zeta) =\mathcal{R}\left(\frac{\xi}{2}\right)\mathcal{R}\left(\frac{\pi}{4}\right)q^{Q}_{q\phi}\mathcal{R}\left(\frac{\rho}{2}\right)(q^{Q}_{q\phi})^{-1}\mathcal{R}\left(-\frac{\pi}{4}\right)\mathcal{R}\left(-\frac{\xi}{2}\right)\mathcal{R}\left(\frac{\xi+\zeta}{2}\right),
\label{euler03}
\end{equation}
\end{widetext}
where $\mathcal{R}(\varphi)$ is $2\times2$ rotation matrix, an element of the SO(2) ($\cong$ U(1) $\subset$ SU(2)) group and is expressed as 
\begin{equation}
    \mathcal{R}(\varphi) = \begin{bmatrix}
        \cos\varphi & -\sin\varphi \\[8pt]
        \sin\varphi & \cos\varphi
    \end{bmatrix}
\end{equation}
and $q^{Q}_{\alpha(\phi)}$ represents the Jones matrix of the $q^{Q}$-plate with fast axis orientation $\alpha(\phi)$. After performing some mathematical algebra, it is straightforward to show that 
\begin{align}
     \mathcal{U}(\xi, \rho, \zeta)& = q^{Q}_{q\phi+\frac{\xi}{2}+\frac{\pi}{4}}q^{H}_{q\phi-\frac{\pi}{4}+\frac{\xi+\rho-\zeta}{4}}q^{Q}_{q\phi+\frac{\pi-2\xi}{4}}
     \label{001} \\
     & = q^{H}_{q\phi-\frac{\pi}{4}+\frac{\xi+\rho-\zeta}{4}}q^{Q}_{q\phi+\frac{\pi}{4}+\frac{\rho-\zeta}{2}}q^{Q}_{q\phi+\frac{\pi}{4}-\frac{\zeta}{2}}
     \label{002} \\
     & = q^{Q}_{q\phi+\frac{\xi}{2}+\frac{\pi}{4}}q^{Q}_{q\phi+\frac{\pi}{4}+\frac{\rho+\zeta}{2}}q^{H}_{q\phi-\frac{\pi}{4}+\frac{\xi+\rho-\zeta}{4}}.
     \label{003}
\end{align}
In the above equations, $q^{H}_{\alpha(\phi)}$ represents the Jones matrix of the $q^{H}$-plate with fast axis orientation $\alpha(\phi)$. Two identities 
\begin{align}
q^{Q}_{q\phi+\alpha_{0}}q^{H}_{q\phi+\widetilde{\alpha}_{0}} &= q^{H}_{q\phi+\widetilde{\alpha}_{0}}q^{Q}_{q\phi+2\widetilde{\alpha}_{0}-\alpha_{0}}, \\
q^{H}_{q\phi+\widetilde{\alpha}_{0}}q^{Q}_{q\phi+\alpha_{0}} &= q^{Q}_{q\phi+2\widetilde{\alpha}_{0}-\alpha_{0}}q^{H}_{q\phi+\widetilde{\alpha}_{0}},
\end{align}
are used in Eq. (\ref{001}) to obtain Eqs. (\ref{002}) and (\ref{003}), respectively. This shows that an SU(2) element with \textit{modified} Euler angle parameterization can be realized as the combination of two $q^{Q}$-plates and one $q^{H}$-plate arranged in any order. Given that SU(2) is a three-parameter continuous Lie group, the minimal set of optical elements required to span the entirety of the SU(2) transformations is comprised of three $q$-plates. Therefore, the specific configurations presented in Eqs. (\ref{001}), (\ref{002}) and (\ref{003}) are optimal. It is evident from these equations that all the $q$-plates involved have the same topological charge $q$, with only their offset angles differing. Hence, these configurations are capable of performing holonomic polarization transformations on HOPS of order $\eta = q$. Furthermore, these configurations span the full SU(2) group with modified Euler angle parameterization. Consequently, the results establish that, under the holonomy condition \cite{umar2025holonomically}, these configurations ($q^{Q}q^{Q}q^{H}$, $q^{Q}q^{H}q^{Q}$ and $q^{H}q^{Q}q^{Q}$) can realize any arbitrary polarization evolutions on the HOPS. In these arrangements, the fast axis orientation of the waveplates is given by $\alpha(\phi) = q\phi + \alpha_{0}(\xi, \rho, \zeta)$. Here, the offset angle dials serve as critical control parameters governing the polarization transformations. For $q=0$, the inhomogeneous waveplate become homogeneous, and the resulting configuration corresponds to the Simon-Mukunda universal SU(2) gadget.\\
\indent
Revisit the Eq. (\ref{psi_01}) where the complex amplitudes are given by
$\psi_{R} = \frac{1}{\sqrt{2}}\left[\cos\chi^{(\eta)} + \sin\chi^{(\eta)}\right] e^{-i\gamma^{(\eta)}}$ and
$\psi_{L} = \frac{1}{\sqrt{2}}\left[\cos\chi^{(\eta)} - \sin\chi^{(\eta)}\right] e^{i\gamma^{(\eta)}}$. Next, consider three $q$-plates, each with identical topological charge $q$ and fast axis orientation $\alpha_{j}(\phi)=q\phi+\alpha_{0j}$ for $j=1$, $2$ and $3$, arranged in the sequence $q^{Q}q^{H}q^{Q}$. When the input HOPS beam, as described in Eq. (\ref{psi_01}), passes through this configuration, the resulting output beam (under the holonomy condition) is expressed as
\begin{align}
|\psi^{'}_\ell\rangle &= \left[J_{\alpha_{3}(\phi)}\cdot J_{\alpha_{2}(\phi)}\cdot J_{\alpha_{1}(\phi)}\right]|\psi_\ell\rangle \nonumber \\
 &= \psi_{R}^{'}\ket{R_{\ell}}+\psi_{L}^{'}\ket{L_{\ell}},
\label{output_hops}
\end{align}
where the complex amplitudes $\psi_{R}^{'}$ and $\psi_{L}^{'}$ is given by
\begin{widetext}
\begin{align}
    \psi_{R}^{'} &= -\frac{1}{\sqrt{2}} \left[( \cos\chi^{(\eta)} + \sin\chi^{(\eta)}) \cos\left( \alpha_{01} - 2\alpha_{02} + \alpha_{03} \right) e^{i(\alpha_{01} - \alpha_{03} - \gamma^{(\eta)})} \right. \nonumber \\
    & \quad \left. + ( \cos\chi^{(\eta)} - \sin\chi^{(\eta)}) \sin\left( \alpha_{01} - 2\alpha_{02} + \alpha_{03} \right) e^{-i(\alpha_{01} + \alpha_{03} - \gamma^{(\eta)})} \right],
\end{align} 
\begin{align}
    \psi_{L}^{'} &= \frac{1}{\sqrt{2}} \left[( \cos\chi^{(\eta)} + \sin\chi^{(\eta)}) \sin\left( \alpha_{01} - 2\alpha_{02} + \alpha_{03} \right) e^{i(\alpha_{01} + \alpha_{03} - \gamma^{(\eta)})} \right. \nonumber \\
    & \quad \left. - ( \cos\chi^{(\eta)} - \sin\chi^{(\eta)}) \cos\left( \alpha_{01} - 2\alpha_{02} + \alpha_{03} \right) e^{-i(\alpha_{01} - \alpha_{03} - \gamma^{(\eta)})} \right].
\end{align} 
\end{widetext}
The Eq. (\ref{output_hops}) shows that the output beam remains on the same HOPS, as it is a superposition of right- and left-circularly polarized vortex beams with opposite topological charges, and the topological charge is the same as that of the input beam. This also holds for the other two configurations, $q^{Q}q^{Q}q^{H}$ and $q^{H}q^{Q}q^{Q}$, as well. Therefore, these combinations serve as a valid SU(2) gadget for the HOPS. In the standard SU(2) gadget for PS, the relative orientation of the fast axes of the involved homogeneous waveplates is pivotal in determining polarization evolution. Similarly, in this case, the relative orientation of the offset angle of the involved $q$-plate fulfills this essential role. \\
\indent
While a system composed of two $q^H$-plates and one $q^{Q}$-plate, regardless of their sequential arrangement ($q^{Q}q^{H}q^{H}$, $q^{H}q^{Q}q^{H}$ and $q^{H}q^{H}q^{Q}$), is again a combination of three $q$-plates, it exhibits a limited capability in realizing the full SU(2) Lie group. This is due to the fact that these setups can be reduced to that of a two $q$-plate configuration. Consequently, these SU(2) elements form a two-parameter subset within the SU(2) group, rather than spanning the entire three-parameter group. In essence, these systems act as constrained SU(2) operators, unable to realize arbitrary rotations in the SO(3) space of HOPS.\\
\indent
The Jones matrices of the $q^Q$-plate and $q^H$-plate are symmetric SU(2) matrices, corresponding to the eighth and fourth roots of the identity element, respectively, i.e., $[q^{Q}_{\alpha(\phi)}]^8=\mathbf{I}$ and $[q^{H}_{\alpha(\phi)}]^4=\mathbf{I}$. These features of the $q$-plates enable the new SU(2) gadget to respect the decomposition theorem established in \cite{simon1990minimal}. According to this theorem, any SU(2) matrix can be expressed as a product of three symmetric SU(2) matrices, where two of them are eighth roots and the third is a fourth root of the identity. However, the theorem also establishes that not all SU(2) matrices can be represented as a product of three symmetric matrices when one of them is an eighth root and the remaining two are fourth roots of the identity. The set of elements that admits such a decomposition constitutes only a two-parameter subset of the SU(2) group.\\
\indent
In conclusion, we have presented an SU(2) gadget, involving three $q$-plates, which can realize all the SU(2) polarization transformations on the HOPS. This gadget consists of, two $q^Q$-plates and one $q^H$-plate, arranged in any order. Under the holonomy condition $\eta = q$, each HOPS admits a corresponding SU(2) gadget. The full coverage of HOPS transformations for any order can be achieved by appropriately setting the relative offset angles of the involved $q$-plates. In the Simon–Mukunda SU(2) gadget, polarization evolution on the Poincaré sphere is dictated by the fast-axis orientations of the constituent waveplates, which are adjusted through mechanical rotation. In our configuration, also, the relative offset angles can be adjusted by mechanically rotating the $q$-plate. However, this approach does not work for the gadget belonging to the index space one, where the $q$-plates have the topological charge $q=1$. In this case, the fast-axis orientation displays radial symmetry, making mechanical adjustment ineffective in changing the relative offsets. The theoretical foundation of the SU(2) gadget has been formulated here, and its experimental implementation could offer profound insights into structured light applications.

MU gratefully acknowledges the institute fellowship from the Indian Institute of Technology (IIT) Delhi. PS also acknowledges the financial assistance from the Science and Engineering Research Board (SERB), India (CRG/2022/001267).

\appendix
\nocite{*}

\bibliography{euler}

\end{document}